 \documentclass[final,5p,times,twocolumn,natbib]{elsarticle}

\usepackage{amssymb}
\usepackage{lipsum}
\usepackage{natbib}
\usepackage{hyperref}
\biboptions{sort&compress}
\journal{Physics Letters B}

\begin{document}

\begin{frontmatter}



\title{SU(3) rigid triaxiality in $^{154}$Sm}


\author[first,second,third]{Chunxiao Zhou}
\ead{zhouchunxiao567@163.com}
\author[first,second,third]{Xue Shang}
\author[forth]{Tao Wang}
\ead{suiyueqiaoqiao@163.com}
\affiliation[first]{organization={College of Mathematics and Physics, Hunan University of Arts and Sciences, Changde 415000, China}
       }
\affiliation[second]{organization={Hunan Province Key Laboratory of Optoelectronic Information Integration and Optical Manufacturing Technology, Changde 415000, China}
}

\affiliation[third]{organization={Hunan Province Higher Education Institution’s Key Laboratory of Information Detection and Intelligent Processing Technology, Changde 415000, China}
}

\affiliation[forth]{organization={College of Physics, Tonghua Normal University, Tonghua 134000, People's Republic of China}
}

\begin{abstract}
The $^{154}$Sm nucleus, which has long been regarded as a typical quantum system exhibiting axial symmetry, has been recently suggested to exhibit a small degree of triaxiality by Otsuka \textit{et al.}. This small triaxiality was recently observed using the $\gamma$ decay of the $^{154}$Sm isovector giant dipole resonance. Thus further studying this small triaxiality is very necessary for understanding the nature of the collectivity in nuclei. In this paper, the rigid triaxiality in $^{154}$Sm is investigated using the newly proposed SU3-IBM theory. If the irreducible representation (irrep) of the ground state is (18,2) in the SU(3) symmetry limit, the new model can effectively reproduce the experimental data of $^{154}$Sm, showing good agreements with realistic energy spectra, B(E2) values, and quadrupole moments.  This result further confirms the vality of the SU3-IBM for the description of the rigid triaxiality in realistic nuclei, and reveals the SU(3) dominance of the collectivity motions in nuclei.
\end{abstract}



\end{frontmatter}




\section{Introduction}
\label{introduction}

The atomic nucleus, as a quantum system of interacting nucleons, exhibits a well-defined surface with a certain geometric shape. In traditional perspective, as Aage Bohr stressed in his Nobel lecture with an example of $^{166}$Er, a large fraction of heavy nuclei exhibit axial symmetry and adopt a prolate ellipsoidal shape \cite{Bohr}. This configuration has long been regarded as the textbook picture of the nuclear shape for the large-deformed nuclei \cite{Bohr1975}. However, the violation of axial symmetry in nuclei has been qualitatively addressed in many investigations, and an alternative picture with rigid triaxial shapes has gained substantial supports from many investigations \cite{davydov1958,davydov1959,smirnov200,wood2004,allmond2010}. The rigid triaxial shapes allow for more complex rotational behaviors and have been extensively investigated, but they were not yet recognized as a major paradigm. Recently some new viewpoints emerged and new studies showed that a large number of heavy deformed nuclei, including $^{166}$Er and $^{154}$Sm, which previously were believed to exhibit axial symmetry, actually demonstrate rigid triaxiality \cite{otsuka2019,tsunoda2023,otsuka2025}. Importantly, this prediction was also observed in recent experiment \cite{Kleeman25}. Thess findings indicate  that rigid triaxiality plays a much more prominent and critical role than previously recognized \cite{otsuka2025,Kleeman25}. 


The interacting boson model (IBM) \cite{Iachello75,iachello1987} provides a suitable theoretical framework for studying various nuclear collective modes and various shape transitions between them  \cite{Warner02,Casten06,Casten07,Bonatsos09,Casten09,Jolie09,Casten10,Jolos21,Fortunato21,Cejnar21,Bonatsos24,Jolie00,Cejnar03,Iachello04,Wang08}. The IBM-1, where the Hamiltonian does not distinguish between protons and neutrons, has U(6) symmetry, and four dynamical symmetry limits are included: the U(5) symmetry limit (spherical shape), the SU(3) symmetry limit (prolate shape), the O(6) symmetry limit ($\gamma$-soft rotation) and the $\overline{\mathrm{SU(3)}}$ symmetry limit (oblate shape) \cite{Jolie01}. It was found that, up to two-body interactions, the IBM-1 can not decribe the rigid triaxial shape \cite{Isacker81}. Thus in the IBM-1, if the rigid triaxility needs to be considered, higher-order interactions must be included, such as the 6-$d$ interaction $[d^{\dag}d^{\dag}d^{\dag}]^{(L)}\cdot[\tilde{d}\tilde{d}\tilde{d}]^{(L)}$ \cite{Isacker84}.

The collective prolate rotation in nuclei can be effectively described by the SU(3) symmetry limit of the IBM-1, where axial symmetry emerges naturally when only up to two-body interactions $-[\hat{Q}\cdot\hat{Q}]^{(0)}$ are considered ($\hat{Q}$ is the SU(3) quadruple operator) \cite{ginocchio1980,dieperink1980}. Recently, it was found that a very tiny region of triaxiality can arise with the inclusion of the cubic $[\hat{Q}\times\hat{Q}\times\hat{Q}]^{(0)}$ term, and this cubic interaction can describe the oblate shape \cite{fortunato2011}, which is different from the $\overline{\mathrm{SU(3)}}$ description \cite{Jolie01}. Rigid triaxial rotation can be realized when up to SU(3) four-body interactions are taken into account \cite{smirnov200}. A quantum rigid rotor can also be realized by emulating the mapping between the rigid triaxial rotor and the SU(3) shell model \cite{leschber1987,castanos1988}. The relationship between the quadrupole deformation variables $(\beta,\gamma)$ and the SU(3) irreducible representation ($\lambda$, $\mu$) can be derived, enabling the description of well-deformed nuclei. Thus interactions with SU(3) symmetry provides a natural framework for describing various rigid triaxial shapes.

\begin{figure}
	\centering 
\includegraphics[width=0.48\textwidth]{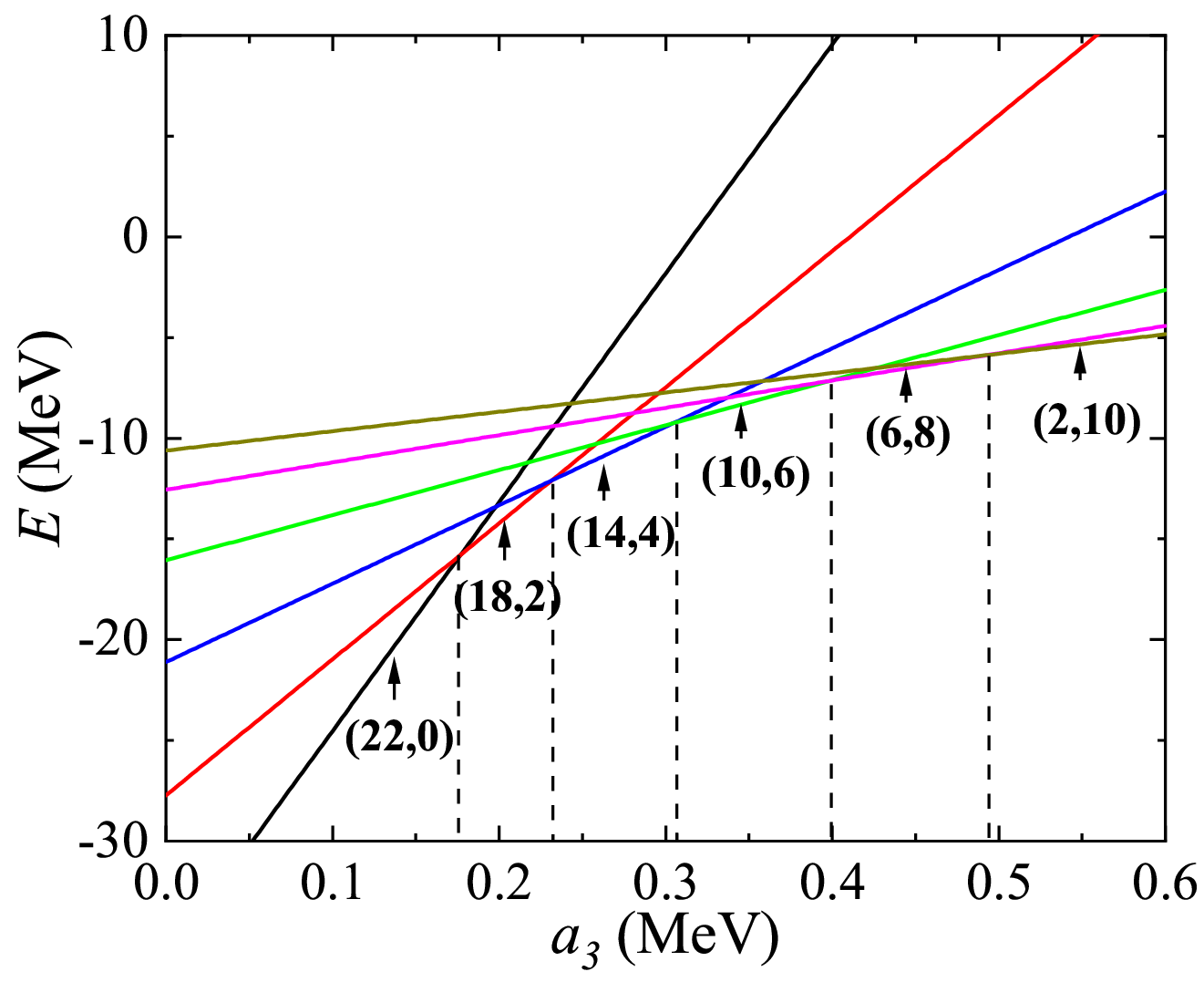}	
	\caption{The low-lying $0^{+}$ levels as a function of $a_{3}$ for $\hat{H}_{S}$ and $N=11$. Other parameters are $a_{1}=1.4516$ MeV, $a_{2}=0.0327$ MeV.} 
	\label{fig1}%
\end{figure}

\begin{figure}
	\centering 
	\includegraphics[width=0.45\textwidth]{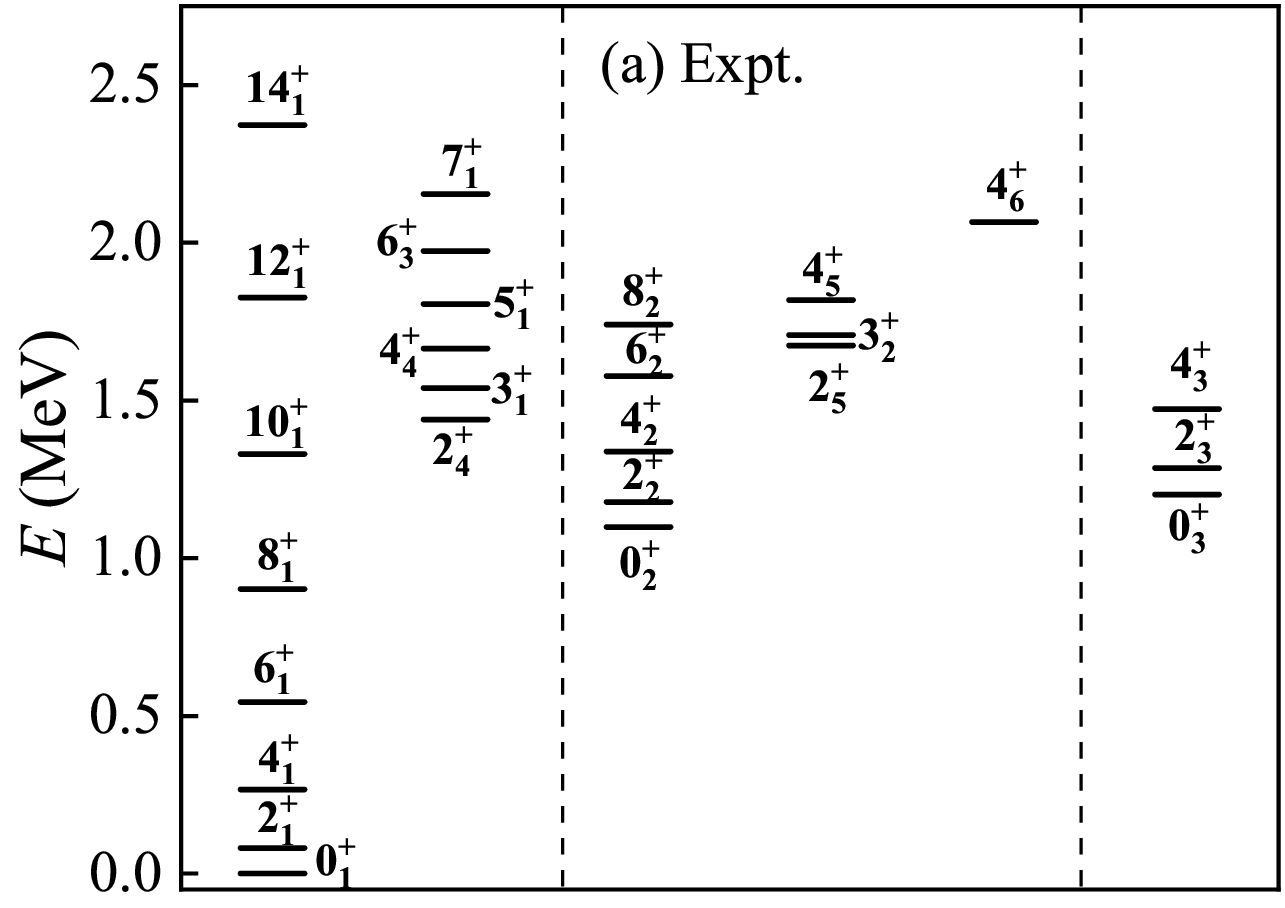}	
        \includegraphics[width=0.45\textwidth]{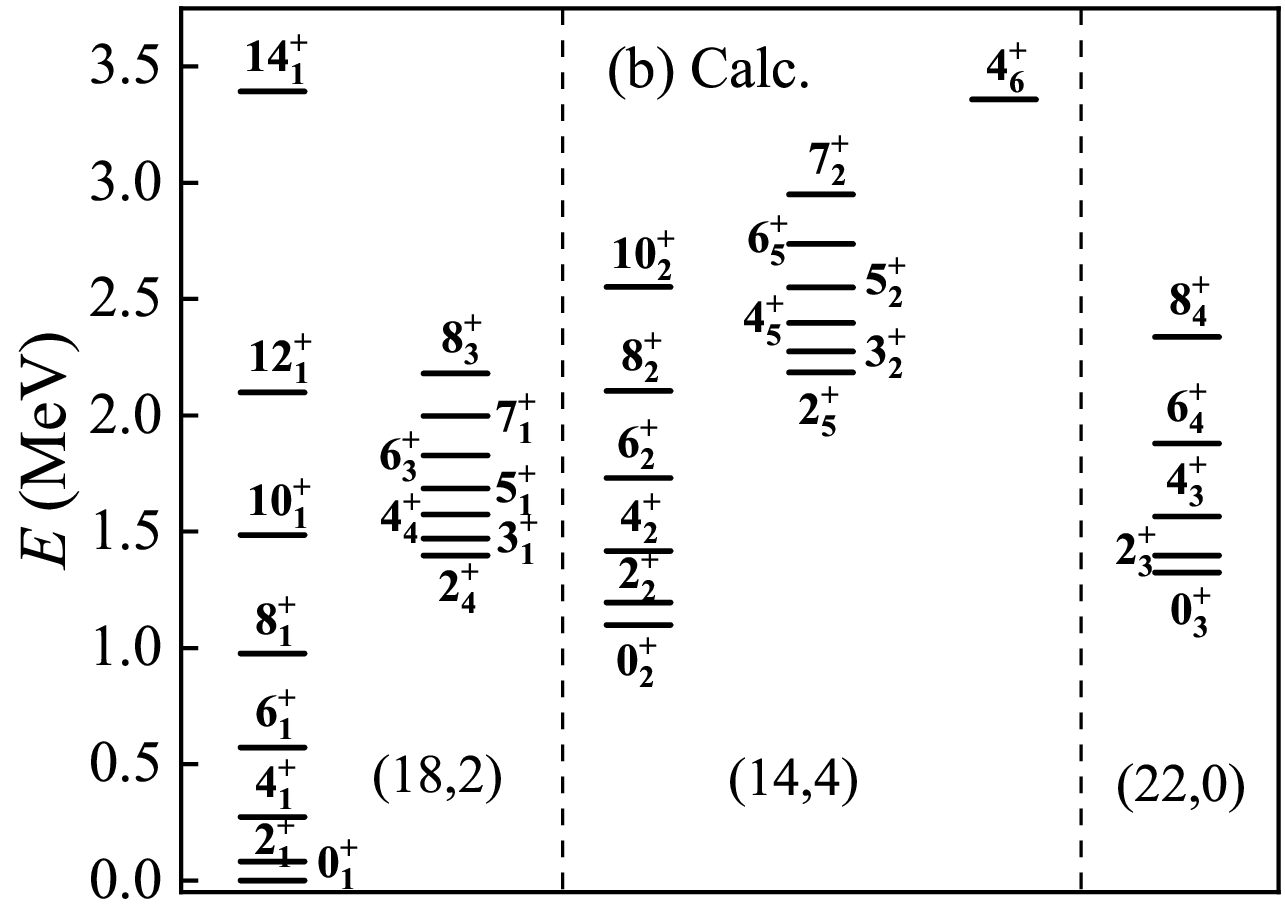}	
	\caption{The experimental and calculated partial low-lying levels in $^{154}$Sm. (a) Experimental data are taken from \cite{smdata}. (b) Calculated results are based on the SU3-IBM hamiltonian (1)  with $\alpha=0.0604$ MeV, $a_{1}=1.4516$ MeV, $a_{2}=0.0327$ MeV, $a_{3}=0.1988$ MeV, $t_{1}=0.002382$ MeV, $t_{2}=-0.002608$ MeV and $t_{3}=0.0484$ MeV.}
 
	\label{fig2}%
\end{figure}

Although the IBM-1 with the 6-$d$ interaction seems reasonable and provides a framework for describing various quadrupole deformations, certain experimental anomalies remain difficult to be exlained, such as the B(E2) anomaly \cite{grahn2016,sayg2017,goasduff2019} and the Cd puzzle \cite{garrett2012,batchelder2012,garrett2019}. Recently, an extension of the IBM with SU(3) higher-order interactions  (SU3-IBM) was proposed, which incorporates both the U(5) symmetry limit and the SU(3) symmetry limit (some biases towards the O(6) symmetry limit may be also needed in some problems). This new model provides an effective description for the collective behaviors of atomic nuclei. It offers novel explanations for the BE(2) anomaly \cite{Wang20,Zhang22,Wangtao,Zhang24,Pan24,Zhang25,Zhang252,Teng25,Cheng25,Li25} and the Cd puzzle \cite{Wang22,Wang25,WangPd106,Zhao25}, and gives a better description for the $\gamma$-soft behaviors in $^{196}$Pt \cite{wang2024}. In addition, it also explains the prolate-oblate shape asymmetric transitions in the Hf-Hg region \cite{wang2023}, describes the E(5)-like spectra observed in $^{82}$Kr \cite{zhou2023}, and reveals the boson number odd-even effect in $^{196-204}$Hg \cite{WangHg}. 

Recently, the $^{154}$Sm nucleus attracts much attentions. Theoretically, Otsuka \textit{et al.} proposed that this nucleus has small triaxiality with $\gamma=3.5^{\circ}$ \cite{otsuka2019,tsunoda2023,otsuka2025}, and even importantly, this small triaxiality was experimentlly confirmed with the $\gamma$ decay of the $^{154}$Sm isovector giant dipole resonance \cite{Kleeman25}. In this experiment, $\gamma$ is 5.0(15)$^{\circ}$, which is somewhat larger than the theoretical prediction. Thus further studying this small triaxiality is very important for understandig the collectivity in large-deformed nuclei. In this paper, the SU(3) rigid triaxiality in $^{154}$Sm is investigated through the SU3-IBM. In the new model, the SU(3) symmetry serves as a central role in describing quadrupole deformations, where varying degrees of triaxiality correspond to different SU(3) irreps. Our calculations demonstrates that the energy spectra, B(E2) transition strengths and quadrupole moments fit well with the experimental results if the SU(3) irreducible representation (irrep)  of the ground state is (18,2) in the SU(3) symmetry limit. Thus the SU(3) symmetry dominates the collective triaxial motion in heavy deformed nuclei. In another paper, similar conclusion can be also obtained with $^{166}$Er \cite{ZhouEr}.

\section{The Hamiltonian}

 To build rigid triaxial modes in the SU3-IBM, the Hamiltonian used in this paper is expressed as

\begin{equation}
\hat{H}=\alpha\hat{n}_{d}+\hat{H}_{Tri},
\label{eq1}
\end{equation}
where $\alpha$ is the fitting parameter, $n_{d}=d^{\dag}\cdot\tilde{d}$ is the $d$-boson number operator in the U(5) symmetry limit. The rigid rotor Hamiltonian $\hat{H}_{
Tri}$ can be divided into static and dynamic parts

\begin{equation}
\hat{H}_{Tri}=\hat{H}_{S}+\hat{H}_{D},
\label{eq2}
\end{equation}
where 
\begin{equation}
\hat{H}_{S}=-\frac{a_{1}}{2N}\hat{C}_{2}[\mathrm{SU(3)}]+\frac{a_{2}}{2N^{2}}\hat{C}_{3}[\mathrm{SU(3)}]+\frac{a_{3}}{2N^{3}}\hat{C}_{2}^{2}[\mathrm{SU(3)}], 
\label{eq3}
\end{equation}
\begin{equation}
\hat{H}_{D}=t_{1}[\hat{L}\times \hat{Q} \times \hat{L}]^{(0)}+t_{2}[(\hat{L}\times \hat{Q})^{(1)} \times (\hat{L} \times \hat{Q})^{(1)}]^{(0)}+t_{3}\hat{L}^{2},
\label{eq4}
\end{equation}
where $a_{1}$, $a_{2}$, $a_{3}$, $t_{1}$, $t_{2}$ and $t_{3}$ are the six fitting parameters. $\hat{L}=\sqrt{10}[d^{\dag}\times\tilde{d}]^{(1)}$, $\hat{Q}=[d^{\dag}\times\tilde{s}+s^{\dag}\times \tilde{d}]^{(2)}-\frac{\sqrt{7}}{2}[d^{\dag}\times \tilde{d}]^{(2)} $ are the SU(3) algebraic generators. The SU(3) Casimir operators are defined as
\begin{equation}
\hat{C}_{2}[\mathrm{SU(3)}]=2\hat{Q}\cdot\hat{Q}+\frac{3}{4}\hat{L}^{2},
\label{eq5}
\end{equation}
\begin{equation}
\hat{C}_{3}[\mathrm{SU(3)}]=-\frac{4\sqrt{35}}{9}[\hat{Q}\times\hat{Q}\times\hat{Q}]^{(0)}-\frac{\sqrt{15}}{2}[\hat{L}\times\hat{Q}\times\hat{L}]^{(0)}.
\label{eq6}
\end{equation}
The eigenvalues of the two SU(3) Casimir operators can be expressed in terms of the SU(3) irreps ($\lambda,\mu$)
\begin{equation}
\langle \hat{C}_{2}[\mathrm{SU(3)}] \rangle=\lambda^{2}+\mu^{2}+3\lambda+3\mu+\lambda\mu,
\label{eq7}
\end{equation}
\begin{equation}
\langle \hat{C}_{3}[\mathrm{SU(3)}] \rangle=\frac{1}{9}(\lambda-\mu)(2\lambda+\mu+3)(\lambda+2\mu+3).
\label{eq8}
\end{equation}
The ground-state energy of the rigid triaxial Hamiltonian (2) is obtained by $E_{g}=\langle \hat{H}_{S}\rangle=f(\lambda,\mu)$ evaluated at the optimal values $(\lambda_{0},\mu_{0})$, which in turn are determined by the parameters $a_{1}$, $a_{2}$, $a_{3}$. The dynamical interactions contained in  $\hat{H}_{D}$ do not contribute to the ground state. 

\begin{table}
\centering 
\setlength{\tabcolsep}{5.0mm}
\begin{tabular}{l c c c} 
 \hline
 \hline
 \quad & Expt. & MCSM & SU3-IBM \\ 
 \hline
 $0_{1}^{+}$ & 5.0(15)$^{\circ}$  & $\approx$3.7$^{\circ}$  & $\approx$7.2$^{\circ}$  \\ 
 $0_{2}^{+}$ & \quad & $\approx$12.6$^{\circ}$  & $\approx$13.9$^{\circ}$   \\ 
 \hline
 \hline
\end{tabular}
\caption{The deformation parameters $\gamma$ in $^{154}$Sm obtained from different approaches and observables. The experimental data is taken from \cite{Kleeman25} and the MSCM data are taken from Ref. \cite{otsuka2025}. The SU3-IBM results are obtained according to Eq. (\ref{eq13}). }
\label{Table1}
\end{table}

In this Hamiltonian, the SU(3) symmetry governs all the quadrupole deformations:  $-\hat{C}_{2}$[SU(3)] generates the prolate shape, and $\hat{C}_{3}$[SU(3)] yields the oblate shape. The combinations of the square of the second-order Casimir operator $-\hat{C}^{2}_{2}$[SU(3)] with $-\hat{C}_{2}$[SU(3)] and $\hat{C}_{3}$[SU(3)] can lead to the emergence of the triaxial shapes \cite{smirnov200}. Furthermore, by combining the $d$ boson number operator with the rigid triaxial shapes (the coupling of $\hat{n}_{d}$ with $\hat{H}_{S}$), various $\gamma$-soft modes different from the traditional O(6) $\gamma$-soft description can emerge \cite{wang2024}. 

In Ref. \cite{otsuka2025} Otsuka \textit{et al.} demonstrated that $^{154}$Sm exhibits a small triaxiality ($\gamma=3.7^{\circ}$), and this conclusion has been further confirmed by experiment very recently \cite{Kleeman25}. Inspired by these results and in view of the relationship between the quadrupole deformation variable $\gamma$ and the SU(3) irrep ($\lambda,\mu$), it would be of interest to employ the SU(3)-IBM to investigate the SU(3) rigid triaxiality in $^{154}$Sm. In the SU(3)-IBM, the ground state configuration is mainly determined by the SU(3) symmetry limit, where the explicit relationship between the parameters $a_{1}, a_{2}, a_{3}$ in Eq. (\ref{eq3}) and the SU(3) irreps ($\lambda_{0},\mu_{0}$) can be derived as follows \cite{zhou2023}.
\begin{equation}
a_{3}=\frac{a_{1}N^{2}}{2g(\lambda_{0},\mu_{0})}-\frac{a_{2}N}{6g(\lambda_{0},\mu_{0})}(3+\lambda_{0}+2\mu_{0} ),
\label{eq9}
\end{equation}
where $g(\lambda_{0},\mu_{0})=\lambda_{0}^{2}+\mu_{0}^{2}+3\lambda_{0}+3\mu_{0}+\lambda_{0}\mu_{0}$. For any arbitrary SU(3) irrep $(\lambda_{0}, \mu_{0})$, it can be the ground state irrep when the parameters in $\hat{H}_{S}$ satisfy the relationship in Eq. (\ref{eq9}). As shown in Fig. \ref{fig1} for $N=11$, different irreps $(\lambda, \mu)$ successively become the ground state as $a_{3}$ increases. This suggests that by adjusting the $a_{3}$ value, we can achieve any quadrupole shape, including the prolate, triaxial, and oblate shape. In this paper any degree of triaxiality, which corresponds to different irrep $(\lambda_{0},\mu_{0})$ can be obtained by selecting the proper parameters in $\hat{H}_{S}$. 

\begin{figure}
	\centering 
	\includegraphics[width=0.45\textwidth]{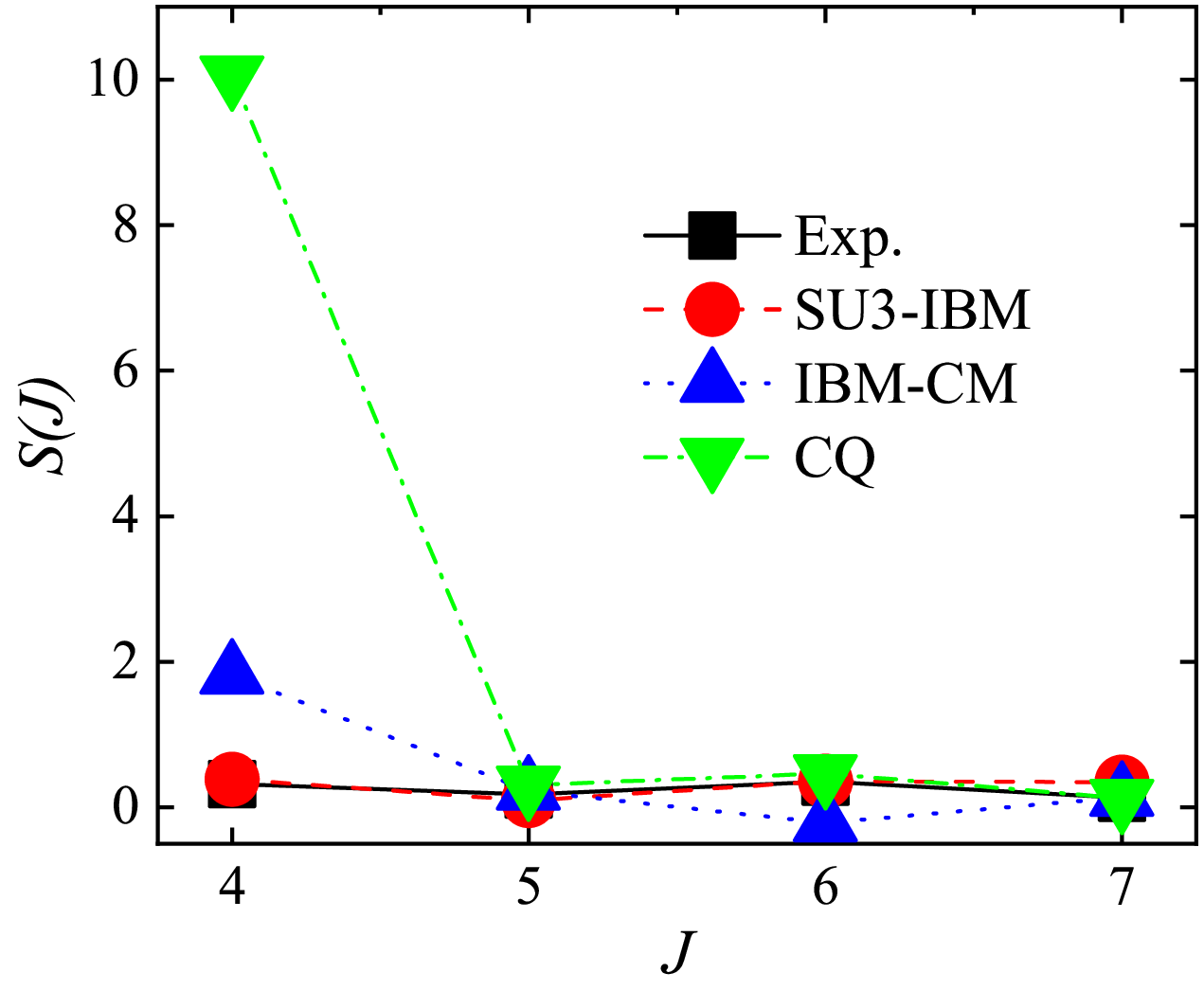}	
	\caption{The staggering parameter $S(J)$ for $^{154}$Sm. The experimental data are taken from \cite{smdata}, and the parameters used in SU3-IBM are same as those used in Fig. (\ref{fig2}). The data of IBM-CM and CQ are obtained from \cite{wu2024}.}
    \label{fig3}
    \end{figure}

\section{Results}

Based on the relationship in Eq. (\ref{eq9}) and the conclusion given in Ref. \cite{otsuka2025}, the irrep $(\lambda,\mu)=(18,2)$ is chosen, and the corresponding parameters in the Hamiltonian are $\alpha=0.0604$ MeV, $a_{1}=1.4516$ MeV, $a_{2}=0.0327$ MeV, $a_{3}=0.1988$ MeV, $t_{1}=0.002382$ MeV, $t_{2}=-0.002608$ MeV and $t_{3}=0.0484$ MeV. Fig. \ref{fig2} (a) and (b) present the comparison between the experimental and calculated energy levels for $^{154}$Sm. The theoretical spectra are organized with the rotational bands governed by the three lowest SU(3) irreps (18,2), (14,4) and (22,0) corresponding to the SU(3) symmetry limit. The levels are arranged in bands characterized by Elliott's quantum number $K$ where, for even $\lambda$ and $\mu$ \cite{smirnov200},

\begin{equation}
    K=\mathrm{min}\lbrace \lambda,\mu \rbrace, \mathrm{min}\lbrace \lambda,\mu \rbrace-2,\dots,2 \quad \mathrm{or}\quad 0,
\end{equation}

\begin{equation}
    L=K,K+1,\dots,\mathrm{max}\lbrace \lambda,\mu \rbrace \quad \mathrm{for} \quad  K\neq0,
\end{equation}

\begin{equation}
    L=0,2 ,\dots,\mathrm{max}\lbrace \lambda,\mu \rbrace \quad \mathrm{for} \quad K=0.
\end{equation}

The experimental low-lying levels shown in Fig. \ref{fig2}(a) can be successfully arranged in a similar way to those in Fig. \ref{fig2}(b). The calculated energy levels are slightly higher than the experiment results, except for the energy band on top of the $2_{4}^{+}$ state. Thus the theoretical calculations nearly perfectly reproduce the experimental results. Some higher levels, such as $8_{3}^{+}$, $10_{2}^{+}$, $6_{5}^{+}$, $7_{2}^{+}$, $6_{4}^{+}$ and $8_{4}^{+}$, are also predicted by our model. More experimental data are excepted to be obtained in the future, which can help verify our theoretical predictions.

\begin{table*}
\centering 
\setlength{\tabcolsep}{5.0mm}
\begin{tabular}{l c c c c c c c} 
 \hline
 \hline
 $L_{i}$ & $L_{f}$ & $\mathrm{Expt.}^{a}$ & SU3-IBM & IBM-CM & CQ & CBS & X(5) \\ 
 \hline
 $2_{1}^{+}$ & $0_{1}^{+}$ & 176(3) & 176 & 176 & 176 & 174 & 176\\ 
 $4_{1}^{+}$ & $2_{1}^{+}$ & 245(6)& 246 & 256& 247& 251 & 278\\ 
 $6_{1}^{+}$ & $4_{1}^{+}$ & 289(8)& 259 & 286& 265& 281 & 348 \\
 $8_{1}^{+}$ & $6_{1}^{+}$ & 319(16)& 255 & 315& 265& 300 & 400 \\
 $10_{1}^{+}$ & $8_{1}^{+}$ & 314(16)& 239 & 340& 255& 314 & \quad \\
 $12_{1}^{+}$ & $10_{1}^{+}$ & 282$^{+19}_{-17}$& 214 &358& 237& \quad & \quad\\
 $0_{2}^{+}$ & $2_{1}^{+}$ & 11.40$^{+28}_{-17}$& 3.26 & 28.1 & 5.2 & 8.4 & \quad\\
 $2_{2}^{+}$ & $0_{1}^{+}$ & 0.32(4)& 0.37 & 0.35 & 0.56 & 0.5 &3.5\\
 $2_{2}^{+}$ & $2_{1}^{+}$ & 0.71(9)& 0.64 &\quad& \quad& 1.4 & 16\\
 $2_{2}^{+}$ & $4_{1}^{+}$ & 1.30$^{+17}_{-14}$& 2.23 &\quad& \quad& 7.1 & 63\\
 $2_{3}^{+}$ & $0_{1}^{+}$ &\quad& 1.49 &2.65& 0.53 & \quad & \quad\\
 $2_{3}^{+}$ & $0_{2}^{+}$ & \quad& 0.97 &0.128 & 1.53 & \quad& \quad\\
 $2_{3}^{+}$ & $2_{1}^{+}$ & \quad& 3.04 &\quad & \quad & \quad& \quad\\
 $2_{3}^{+}$ & $4_{1}^{+}$ & \quad& 0.36 &0.128 & 1.53 & \quad& \quad\\
 $2_{4}^{+}$ & $0_{1}^{+}$ & 1.93(14) & 1.25 & 0.418 & 0.001 & \quad& \quad\\
 $2_{4}^{+}$ & $2_{1}^{+}$ & 3.19$^{+23}_{-27}$ & 1.06 &\quad & \quad & \quad& \quad\\
 $2_{4}^{+}$ & $4_{1}^{+}$ & 0.36(4)& 0.12 &0.178 & 0.015 & \quad& \quad\\
 $4_{2}^{+}$ & $2_{1}^{+}$ & 0.32$^{+14}_{-8}$ & 0.43 & \quad & \quad & 0.1 & 1.8\\
 $4_{2}^{+}$ & $4_{1}^{+}$ & 0.58$^{+31}_{-16}$ & 0.43 &\quad & \quad & 1.1 & 11\\
 $4_{2}^{+}$ & $6_{1}^{+}$ & 0.64$^{+30}_{-16}$ & 2.34 & \quad & \quad & \quad &49\\
 \hline
  \hline
\end{tabular}
\caption{Absolute $B(E2)$ values in W.u. for $E2$ transitions between the low-lying states in $^{154}$Sm. Experimental values are taken from \cite{smdata}. The parameters used in our calculations are same as those used in Fig. (\ref{fig2}). The corresponding effective charge in calculations is $e$=2.06916 (W.u)$^{1/2}$. The data of IBM-CM and CQ are taken from \cite{wu2024}, the data of CBS are taken from \cite{pietralla2004}, the data of X(5) are taken from \cite{moller2012}}
\label{Table1}
\end{table*}

\begin{table}
\centering 
\begin{tabular}{l c c c c} 
 \hline
  \hline
 \quad & Expt. & SU3-IBM & IBM-CM & CQ  \\ 
 \hline
 $Q(2_{1}^{+})$ & $-$1.87(4)  & $-$1.88 & $-$2.34 & $-$2.44\\ 
 $Q(4_{1}^{+})$ & $-$2.2(8)& $-$2.4 & $-$2.9 & $-$3.2 \\ 
 \hline
 \hline
\end{tabular}
\caption{Quadrupole moments in eb for some low-lying states in $^{154}$Sm and various models. The experimental data are taken from \cite{smdata}, and the parameters used in SU3-IBM are same as those used in Fig. (\ref{fig2}). The data of IBM-CQ and CQ are taken from Ref. \cite{wu2024}. }
\label{Table1}
\end{table}

We should note that good descriptions for $^{154}$Sm have also been reported in Ref. \cite{otsuka2025}, but no such many levels. There are two main differences between our calculation and those in Ref. \cite{otsuka2025}. Firstly, small triaxiality of $\gamma\approx3.7^{\circ}$ is suggested in Ref. \cite{otsuka2025}. However, according to the relationship between the SU(3) irrep $(\lambda,\mu)$ and the $\gamma$ quadrupole variable  
\begin{equation}
\gamma=\mathrm{tan}^{-1}\left[\frac{\sqrt{3}(\mu+1)}{2\lambda+\mu+3}\right], 
\label{eq13}
\end{equation}
our model yields a bigger value $\gamma\approx7.2^{\circ}$ based on the SU(3) irrep (18,2), which is larger than the one in \cite{otsuka2025}. The $\gamma$ values for the $0_{1}^{+}$ and $0_{2}^{+}$ states obtained from different approaches are presented in Table 1. For the $0_{1}^{+}$ state, the experimental value is between the two theoretical values of the MCSM and the SU3-IBM. For the $0_{2}^{+}$ state, the results given by the MCSM and the SU3-IBM (corresponding to irrep (14,4)) are very close. Thus our fits qualitatively reproduce the results given in \cite{otsuka2025}, and reveals the validity of the SU(3) dominance on the rigid triaxiality.

Secondly, in Ref. \cite{otsuka2025} there exists a very high $2^{+}$ state near the ground band, which does not appear in our calculation. This is due to the smaller $\gamma$ value than our calculated value. Due to the few energy bands provided in Ref. \cite{otsuka2025}, a more detailed comparison is impossible now. Therefore, we hope that more energy levels can be obtained through state-of-the-art configuration interaction calculations in future, which would allow for a deeper exploration for $^{154}$Sm and a more comprehensive comparison between the two theoretical approaches.

Odd-even staggering in $\gamma$ bands is examined by the quantity \cite{zamfir1991,mccutchan2007} 
\begin{equation}
S(J)=\frac{(E_{J}-E_{J-1})-(E_{J-1}-E_{J-2})}{E_{2_{1}^{+}}},
\end{equation}
where $E_{J}$ denotes the energy of spin-$J$ member of the $\gamma$ band. Because $S(J)$ takes the form of a derivative, it is very sensitive to the distribution of the energy levels. The odd-even staggering for $^{154}$Sm is illustrated in Fig. (\ref{fig3}),  where the experimental data are compared with results from different theoretical models. The black square in Fig. (\ref{fig3}) represents the experimental $S(J)$, which are all small positive values. The results given by CQ \cite{warner1983,wu2024} exhibit very large deviations from the experimental result for $S(4)$. The IBM-CQ method \cite{wu2024,harder1997} also exhibits discrepancies compared to the experimental results, giving a larger value for $S(4)$ and a negative value for $S(6)$. It is evident that the results obtained by our model fit very well with the experimental result, demonstrating a notable improvement over both the CQ and IBM-CM models.

The reduced transitional rate serves as a critical probe for the collective behaviors. The E(2) operator is defined as 
\begin{equation}
 \hat{T}(E2)=e\hat{Q},
\end{equation}
where $e$ is the boson effective charge and $\hat{Q}$ is the same as that involved in Hamiltonian (\ref{eq2}). Once the model parameters are determined, the B(E2) values between the low-lying states in $^{154}$Sm are calculated accordingly, and the results are shown in Table 2. The experimental data and our calculated results are listed in the third and fourth columns, respectively. The last four columns display the results obtained through IBM-CM, CQ, CBS and X(5), respectively. When compared to the experimental results, our calculated results are generally in good agreement with the experimental results, except for the values of $B(E2;0_{2}^{+}\rightarrow2_{1}^{+})$ and $B(E2;4_{2}^{+}\rightarrow6_{1}^{+})$.
 Additionally, it can be seen that the calculated values of $B(E2;2_{2}^{+}\rightarrow4_{1}^{+})$, $B(E2;4_{2}^{+}\rightarrow2_{1}^{+})$ and $B(E2;4_{2}^{+}\rightarrow4_{1}^{+})$ in our model demonstrate notable improvements over the corresponding results obtained using the CBS and X(5) formalisms \cite{pietralla2004,moller2012}. Similarly, the calculated value of $B(E2;2_{4}^{+}\rightarrow0_{1}^{+})$ in our model is more consistent with the experimental result compared to that obtained by the IBM-CM and CQ formalism. 

The quadrupole moment, which serves as a direct indicator of nuclear deformation, offers valuable insights into phenomena such as shape coexistence and triaxiality. Table 3 presents the experimental data and theoretical predictions from various models. It can be seen that calculated results of both the IBM-CM and the CQ deviate significantly from the experimental results. 
In contrast, the results obtained from our model demonstrate remarkable agreement with the experimental results. However, experimental data for the quadrupole moments are scarce, and more experimental values are expected in the future to verify our model. 
Based on the above study, it is evident that the SU3-IBM is capable of simultaneously describing the energy levels, BE(2) values and collective quadrupole moments.

\section{Discussions}

In previous IBM-1, if the higher-order interactions are not considered, the rigid triaxiality can not be described \cite{Isacker81}. If the large deformed nuclei are in fact rigid triaxial \cite{otsuka2025}, the higher-order interactions must be necessary. The experimental confirmation of the small triaxiality in $^{154}$Sm decisively changed the opinions on the large deformed nuclei. Thus in the IBM, higher-order interactions are needed. In our fits, we show that the SU3-IBM provides a self-consistent framework for discussing any rigid triaxial shape. 

In another paper on $^{166}$Er \cite{ZhouEr}, the same conclusion is also obtained. These results reveal that rigid triaxial shapes are universal deformation modes for the large deformed nuclei. This result is exactly what Davydov had speculated before \cite{davydov1958,davydov1959} and was recently reconfirmed by Otsuka \textit{et al.} \cite{otsuka2019,tsunoda2023,otsuka2025}. Thus the opinions on the shape evolution in nuclei are changed for ever.

If studying the rigid triaxial shapes, not only the SU(3) third-order interactions, but also the SU(3) fourth-order interactions are needed. Thus inclusion of other higher-order interactions, such as the 6-$d$ interaction \cite{Isacker84}, can not realize the same fitting effect. In this study, the validity and correctness of the SU3-IBM is determined in a nearly unique way.

\section{Summary and Conclusions}
 In summary, the small triaxiality in the low-lying collective bands in $^{154}$Sm is investigated using the SU3-IBM. In this framework, the quadrupole deformation is governed by the SU(3) symmetry and the shape can be described by a specific SU(3) irrep ($\lambda,\mu$). The energy levels, B(E2) values, and quadrupole moments of $^{154}$Sm are calculated when the SU(3) irrep of the ground state is (18,2) in the SU(3) symemtry limit. The calculated results show good agreement with the experimental data, and qualitatively reproduce the results gevin by Otsuka \textit{et al.}, demonstrating the effectiveness of the SU3-IBM and confirms the SU(3) dominance of the rigid triaxiality in $^{154}$Sm.

\section*{Acknowledgements}

This work is supported by the National Natural Science Foundation of China (Grant No. 12104150), Scientific Research Fund of Hunan Provincial Education Department (Grant No. 24B0639) and the Research Foundation of Hunan University of Art and Science (Grant No. 23ZZ04).








\end{document}